# Designing an intelligent computer game for predicting dysgraphia

Zahra Nevisi, Maryam Tahmasbi

Shahid Beheshti university, computer and data science department

## Abstract

Dysgraphia is a key cognitive disorder impacting writing skills. Current tests often identify dysgraphia after writing issues emerge. This paper presents a set of computer games and uses machine learning to analyze the results, predicting if a child is at risk. The games focus on cognitive differences like visual attention between dysgraphic and typical children. The machine learning model forecasts dysgraphia by observing how kids interact with these games. We also create an algorithm to detect unsuitable testing conditions, acting as a preprocess to avoid mislabeling them as dysgraphia. We developed a machine learning model capable of predicting dysgraphia with 93.24% accuracy in a test group of 74 participants.

## Keywords

Dysgraphia; Human computer interaction; Serious games; Prediction

## 1. Introduction

Dysgraphia is a disorder affecting the written expression of symbols and words. In contemporary culture, we still heavily depend on our ability to communicate using written language; therefore, dysgraphia can be a serious problem. In addition, dysgraphia in a school setting can affect the child's normal development and self-esteem, as well as academic achievements [2].

Developmental dysgraphia is a disorder of writing/spelling skills, closely related to developmental dyslexia. For developmental dyslexia, profiles with a focus on phonological, attentional, visual or auditory deficits have recently been established. Unlike for developmental dyslexia, however, there are only few studies about dysgraphia, in particular about the variability of its causes [1].

Specific reading disorder (dyslexia) and specific writing disorder (dysgraphia), in recent literature categorized as specific learning disabilities (SLD), manifest when starting school, and persist throughout schooling, affecting academic achievement and other spheres of functioning. The frequency of SLD ranges from 5 to 20% with a tendency to increase. [3].

Conventional dysgraphia diagnosis often comes late, after writing struggles are noticed. However, early detection is crucial for effective intervention. Timely recognition can prevent negative outcomes like academic failure, low self-esteem, depression, and social issues. Prompt diagnosis is key to addressing dysgraphia effectively.

This study developed a computer game to detect dysgraphia based on cognitive differences, enabling pre-school diagnosis. The game-based approach offers a stress-free, accessible, and cost-effective alternative to traditional clinics for assessing children's writing disorders.

## 2. literature overview

Table 1 provides an overview of some previous activities.

| Tool Name | Age/Class | Task(s) | ref |
|---|---|---|---|
| BHK: Brave Handwriting Kinder | 1st to 5th grade | Copy | 13 |
| BHK Ado: Rapid Writing Evaluation Scale for Adolescents | 6th to 9th grade | Copy | 14 |
| BVSCO-3: Test for the Evaluation of Writing and Orthographic Ability, 3rd ed. | 6–14 y | Copy Dictation<br>Spontaneous production | 15 |
| CHES: Children's Handwriting Evaluation Scale | 3rd to 8th grade | Copy | 16 |
| CHES-M: Children's Handwriting Evaluation Scale—Manuscript Writing | 1st to 2nd grade | Copy | 17 |
| DASH: Detailed Assessment of Speed of Handwriting | 9–16 y | Alphabet copy at normal and high speed<br>Spontaneous production | 18 |
| DRHP: Diagnosis and Remediation of Handwriting Problems | From 3rd grade | Spontaneous production from images observation | 19 |
| ETCH-M: Evaluation Tool of Children's Handwriting—Manuscript | 1st to 2nd grade | Copy<br>Dictation<br>Spontaneous production<br>Handwriting from memory | 20 |
| EVEDP: Evaluation de la Vitesse d'Ecriture—Dictée Progressive | 2nd to 5th grade | Dictation | 21 |
| HLS: Handwriting Legibility Scale | 9–14 y | Spontaneous production | 22 |
| MMHAP: Mac Master Handwriting Assessment Protocol | Preschool to 6th grade | Copy<br>Dictation<br>Spontaneous production<br>Handwriting from memory | 23 |
| MHA: Minnesota Handwriting Assessment | 1st to 2nd grade | Alphabet<br>Copy | 24, 25 |
| QNST-3 Revised: Quick Neurological Screening Test, 3rd ed. Revised | 5–80 y | Copy | 26 |
| SCRIPT: Scale of Children's Readiness in Printing | N.A. | Copy | 27 |
| TOLH: Test of Legible Handwriting | 2nd to 12th grade | Spontaneous production<br>Text composition at school | 28 |
| THS-R | 6–18 y | Alphabet<br>Copy | 29 |

*Table 1: List of the most commonly used tools for the diagnosis of dysgraphia in children based on the analysis of the handwriting product[4]*

## 3. Approach

To design a dysgraphia prediction game, we first need to identify cognitive differences between dysgraphic and normal cases.

In [1] different tests were carried out with 3rd and 4th grade school children to assess their spelling abilities, tapping into phonological processing, auditory sound discrimination, visual attention and visual magnocellular functions as well as reading. A group of 45 children with developmental dysgraphia was compared to a control group. The results showed that besides phonological processing abilities, auditory skills and visual magnocellular functions affected spelling ability, too. Consequently, by means of a two-step cluster analysis, the group of dysgraphic children could be split into two distinct clusters, one with auditory deficits and the other with deficits in visual magnocellular functions. Visual attention was also related to spelling disabilities, but had no characteristic distinguishing effect for the two clusters.

Many of the theories regarding mechanisms of dysgraphia have been derived from studies of individuals with acquired dysgraphia [8,9]. Writing has been shown to be a complex process that requires the higher order cognition (language, verbal working memory and organization) coordinated with motor planning and execution to constitute the functional writing system [5]. Different writing tasks require different cognitive processes, and individuals with dysgraphia may have disorders in one or more areas. For example, when asked to spell a dictated word, the listener must utilize phonological awareness to access phonological long-term memory and the associated lexical-semantic representations. This in turn activates the orthographic long-term memory to create abstract letter representations that require motor planning and coordination to execute the task of writing, all maintained in the working memory. Spelling a pseudoword or novel word requires the function of sub lexical spelling process that applies known phoneme-graphene conventions to predict the correct spelling. Generating a new word spontaneously would first require the usage of orthographic skills, which would then access the lexical representation. Writing rapidly and fluidly requires motor planning and coordination mediated by the cerebellum.

Throughout the writing task, visual and auditory processing and attention is crucial to the production of legible writing [5].

From the moment the child is born, learning becomes meaningful and it is interpreted as a result of the experiences first in the family and then in school. However, it is sometimes not possible to talk about the fact that learning takes place in all children although the process has taken place in this direction. Sometimes the individual differences that exist in children and the inability to get the necessary support in structuring their learning experiences can be effective in the failure of learning, while sometimes the type of congenital difficulty can be effective. One of these types of difficulty is a specific learning difficulty. It is not always possible for children with specific learning difficulties to learn, even if they do not have any mental problems. In this case, many factors can be effective, especially the problems that children experience in their visual perception can become effective. Since visual perception is the processing of symbols received from the environment in the brain, the problem that may be experienced in this process can also make it difficult to learn this situation [6].

Visual attention inspired the game design, based on theories about cognitive differences in dysgraphic children. The games weren't meant to measure specific cognitive factors, but rather to explore if cognitive skills could predict dysgraphia.

A typical visual scene is complex and filled with a huge amount of perceptual information. The term, “visual attention” describes a set of mechanisms that limit some processing to a subset of incoming stimuli. Attentional mechanisms shape what we see and what we can act. They allow for concurrent selection of some (preferably, relevant) information and inhibition of other information. This selection permits the reduction of complexity and informational overload. Selection can be determined both by the “bottom-up” saliency of information from the environment and by the “top-down” state and goals of the perceiver. Attentional effects can take the form of modulating or enhancing the selected information. A central role for selective attention is to enable the “binding” of selected information into unified and coherent representations of objects in the outside world [7].

## 4. designing and developing the prediction system

This system consists of the following parts: (1) data collection game and unsuitable conditions detection system, (2) data collection game, data augmentation algorithm, dysgraphia detection system. In the following, each of the sections are explained.

### 3.1 Inappropriate conditions detection sub system

temporary confusion can affect cognitive test results in several ways:

- Reduced Attention: Temporary confusion can lead to reduced attention and concentration, resulting in lower performance on cognitive tests that require sustained focus.
- Impaired Memory: Temporary confusion may impair memory function and the ability to recall information, affecting performance on cognitive tests that assess memory recall.
- Decreased Problem-Solving Ability: Temporary confusion can diminish problem-solving abilities, making it harder to effectively tackle complex tasks or questions on cognitive tests.
- Slowed Processing Speed: Temporary confusion can slow down processing speed in the brain, leading to slower response times and potentially lower scores on cognitive tests that assess processing speed.

Based on [10], The confusion positively impacts performance in the memory, but not about attention.

Generally, temporary confusion can temporarily diminish cognitive performance. Fatigue, physical pain, and various other factors can also have the same effect.

As our objective in this research is to evaluate dysgraphia, we recognize that temporary confusion and other external variables can impact test outcomes. Therefore, we suggest incorporating a mechanism for identifying improper conditions to minimize errors in the system.

Game 1 only displays one image per page, featuring simple geometric shapes. Using the left and right arrow keys on the keyboard, the child determines if the image on the current page matches the previous one. The child then proceeds to the next page to continue the game. (Figure 1)

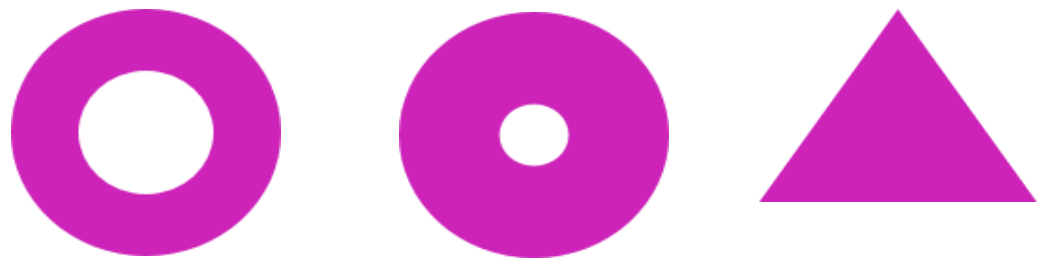

*Figure1 : 3 consecutive frames of game 1*

The effect of color blindness on the visual capabilities is different from dysgraphia. To eliminate the effect of color blindness on the result of this game, all images are monochrome.

The number of frames in this game plays a crucial role. The use of monochrome images might make the game less engaging for children and potentially affect the accuracy of their responses. For examples the game should not be so long that playing the game itself leads to boredom of the children. To address this, the number of images is carefully chosen to ensure that there is no significant difference in a child's error rate between the first and second halves of the frames. As a result, based on the evaluation of children's errors in the game, the number of game pages was kept short enough to avoid any noticeable discrepancy in a child's errors between the initial and final sections of the game.

The resulting record of game 1 is an array as below:

(response time1, result1(true or false)), (response time2, result2(true or false)), …

The sequence of difference between consecutive response times is called *click time seq*.

This game is played by 32 children, 15 confused and 17 normal. Then we calculated the mean and standard deviation of each group, $\mu_{nor}, \sigma_{nor}$ and $\mu_{con}, \sigma_{con}$, respectively. To detect Inappropriate conditions in a child, he plays game1 and calculate the mean $\mu$ of click time seq. The Inappropriate conditions label is set according to the following conditions:

- If $|\mu - \mu_{con}| < \sigma_{con}$ or $\mu < \mu_{con}$, then the child is labeled as confused.
- If $|\mu - \mu_{nor}| < \sigma_{nor}$ or $\mu > \mu_{nor}$, then the child is labeled as normal.
- Otherwise, the child is labeled as confused if the click time seq is decreasing.

The algorithm mentioned above was tested on a sample of 20 children, consisting of 9 confused and 11 normal. The algorithm successfully identified 16 cases, including 6 confused and 10 normal children. The results indicate that implementing this algorithm at the outset of cognitive state diagnosis systems can effectively reduce system errors.

### 3.2 dysgraphia prediction sub system

When it comes to selective attention, the game involves a screen with objects that move around. There are specific objects that the player needs to focus on, and they must track these objects until they move out of the game area and then indicate where they exited.

In the initial stage of the game, the player must track a specific starfish among others that are either moving or stationary, and then click on its destination spot on the table. As the game progresses, the number of target objects increases to three.

In this section, we will discuss the initial stages of the game, which are designed to determine whether the child has noticed the game. It is important to note that the scores obtained in these sections are not included in the final calculations.

In the first section, the game begins with only one target object displayed on the screen.

In the next section, a target object is accompanied by additional objects.

In the next section, there are two target objects and one additional object.

As the game progresses, the number of additional objects gradually increases.

Figure 2 is an image of the game with orange lines added. The orange lines are a sample that show the path that the children had to follow the starfish to find its exit point.

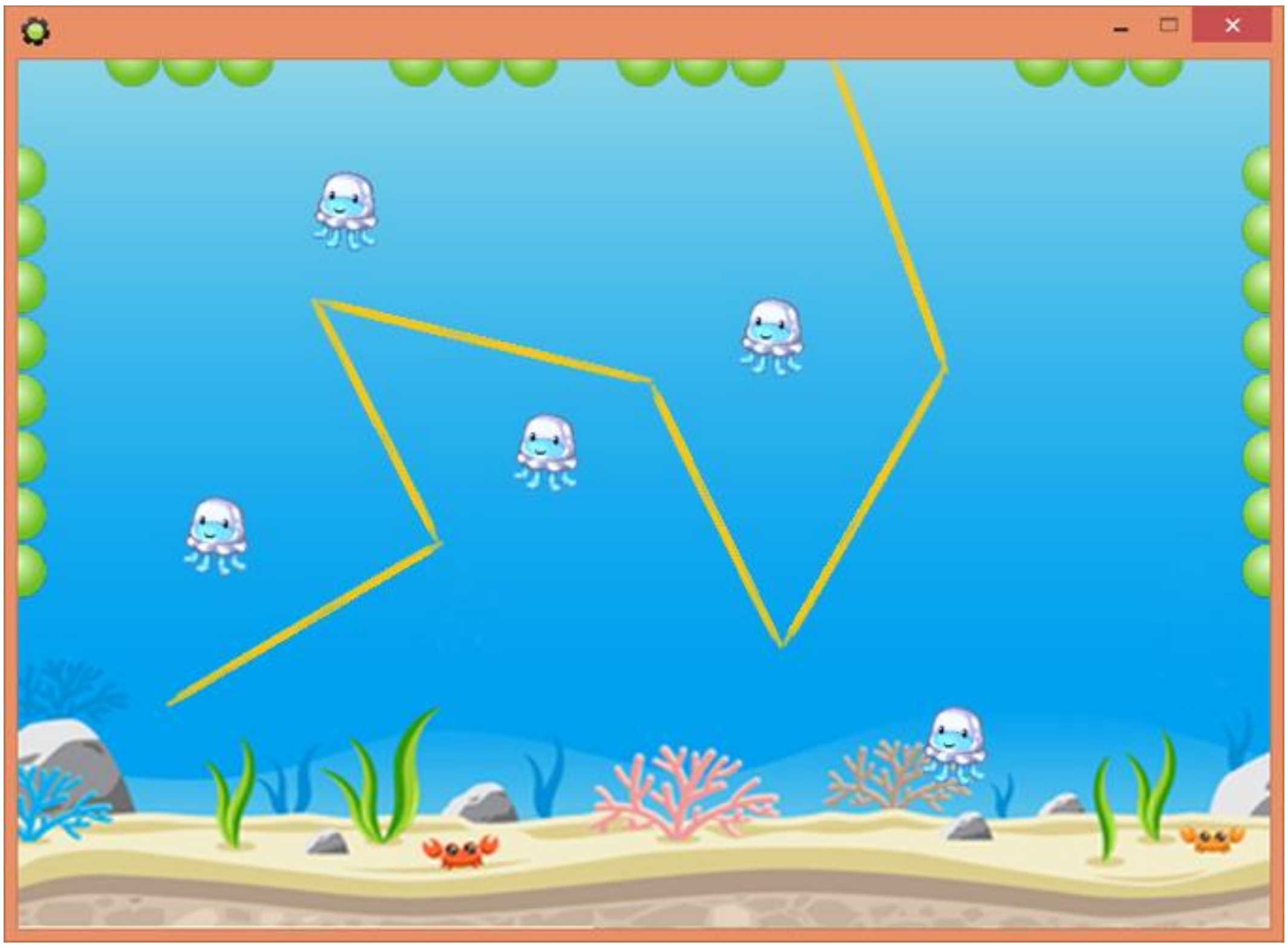

*Figure 2   An image of the game with orange lines added.*

In the sequel to the game, players must track a comet in the sky among the surrounding fixed and moving stars and other comets.

In part three, the starfish becomes the object in motion.

The resulting record of game 2 that used for training and testing the system is an array as below:

(final response time, final result).

In the modeling phase, the resulting record of game 2 that used for data augmentation, is an array as below:

(response time1, result1(true or false)), (response time2, result2(true or false)), …

This type of record is gathered only for using in data augmentation as described in next section.

In the case where we have more than one target, if there is no reaction recorded from the time one target leaves the screen until the next target leaves, the next target's exit time is recorded as the response time and the error result is recorded as the response quality.

Following the design and modification of games, a study sample comprising 45 normal and 30 dysgraphic Persian-speaking children was selected through random sampling. The participants were chosen from a pool of children whose parents consented to their involvement and who themselves expressed willingness to participate in the study. During the training phase, to ensure participant confidentiality, children's game performance data was collected without any personally identifiable information. This approach safeguarded the children's privacy while allowing for training the system. Prior to the commencement of the study, the homogeneity of the sample groups was assessed in terms of Intelligence Quotient (IQ), ensuring a standardized baseline for comparison. This rigorous selection and assessment process aimed to provide a solid foundation for investigating the potential benefits of specially designed games in supporting children with and without dysgraphia.

Class imbalance (CI) is a challenging problem for machine learning that arises with a disproportionate ratio of instances. CI problem is a typical problem in classification tasks [11] in a wide range of applications. In the case of binary classification, the larger number of instances makes the majority class, while a much lower number of instances makes the minority class.

Machine learning models used for classification are typically designed with the assumption of an equal number of instances for each class [12].

In CI datasets, machine learning models tend to be more biased towards the majority class, causing improper classification of the minority class and leading to poor classification performance. This causes high true positive rates (TPR) but a low true negative rate (TNR) when most instances are positive [12].

In this study, the dysgraphia class, which is the main problem for diagnosis, initially has fewer members. We used data augmentation to balance the dysgraphia class.
The method used is as follows: three records are selected from the dysgraphia class and their first, middle and last thirds are randomly selected and a new record is created by connecting to each other. At the connection points, in order to preserve the ascending component of the response time, if necessary, the times will have a displacement by maintaining the time intervals. The three selected records are discarded and the remaining records are used to create the next added record.

In order to generate new records, select 3 records randomly: R1, R2, R3

Split each one to 3 parts: R(i,1), R(i,2), R(i,3). For i=1,2,3.

Generate a record R=( R(new,1), R(new,2), R(new,3) with R(new,j) randomly selected from { R(i,1), R(i,2), R(i,3)} For i=1,2,3.

The new record is valid if the time sequence is ascending.

For each game output data, we computed an array of the following: Total time, total score, time of the first wrong reaction, time of the last correct reaction.

This array, which provides a summary of long game output records, is further used to train and test the system.

To design an intelligent detection system, the C4.5 algorithm was used.

The C4.5 algorithm was developed by Ross Quinlan as an enhancement of the ID3 algorithm. It is used to generate a decision acyclic graph (tree) that can be utilized for classification tasks in machine learning and data mining.

Steps of the C4.5 Algorithm are as below:

Check for Base Cases:

If all the instances belong to a single class, create a leaf node with that class.

If no attributes are left, create a leaf node with the most frequent class.

If the dataset is empty, return a leaf node with the default class.

Calculate Information Gain Ratio:

For each attribute, compute the information gain ratio.

Select the Best Attribute:

Choose the attribute with the highest information gain ratio to split the data 1.

Create a Decision Node:

Generate a decision node and split the dataset into subsets based on the selected attribute.

Recursively Create Subtrees:

Apply the algorithm recursively to the subsets, creating subtrees for each branch.

Pruning:

Perform pruning to remove branches that do not add significant predictive power, thus avoiding overfitting.

## 5. Result, discussion and conclusion

In the test phase, to maintain confidentiality, we did not store any additional data except the children's actual label and the result label.

The system was tested with 74 children, as showed in Table 1.

| | |
|---|---|
| 74 | The total number of participants |
| 53 | The number of children labeled as normal |
| 21 | Number of children with dysgraphic label |

*Table 2 participants*

Three cases were not in a suitable condition for testing dysgraphia by the Inappropriate condition detection section, as showed in Table 3.

| | |
|---|---|
| 3 | The number of children whose conditions were found to be unsuitable for testing by the system |
| 2 | Number of children labeled normal |
| 1 | Number of children with dysgraphic label |

*Table 3 Diagnosing the appropriateness of the child's conditions for measuring dysgraphia*

Of the remaining 71 children, six cases were wrongly labeled and 65 cases were correctly labeled by the system as showed in Table 4. Based on this test, the overall accuracy of the system was estimated at 93.24.

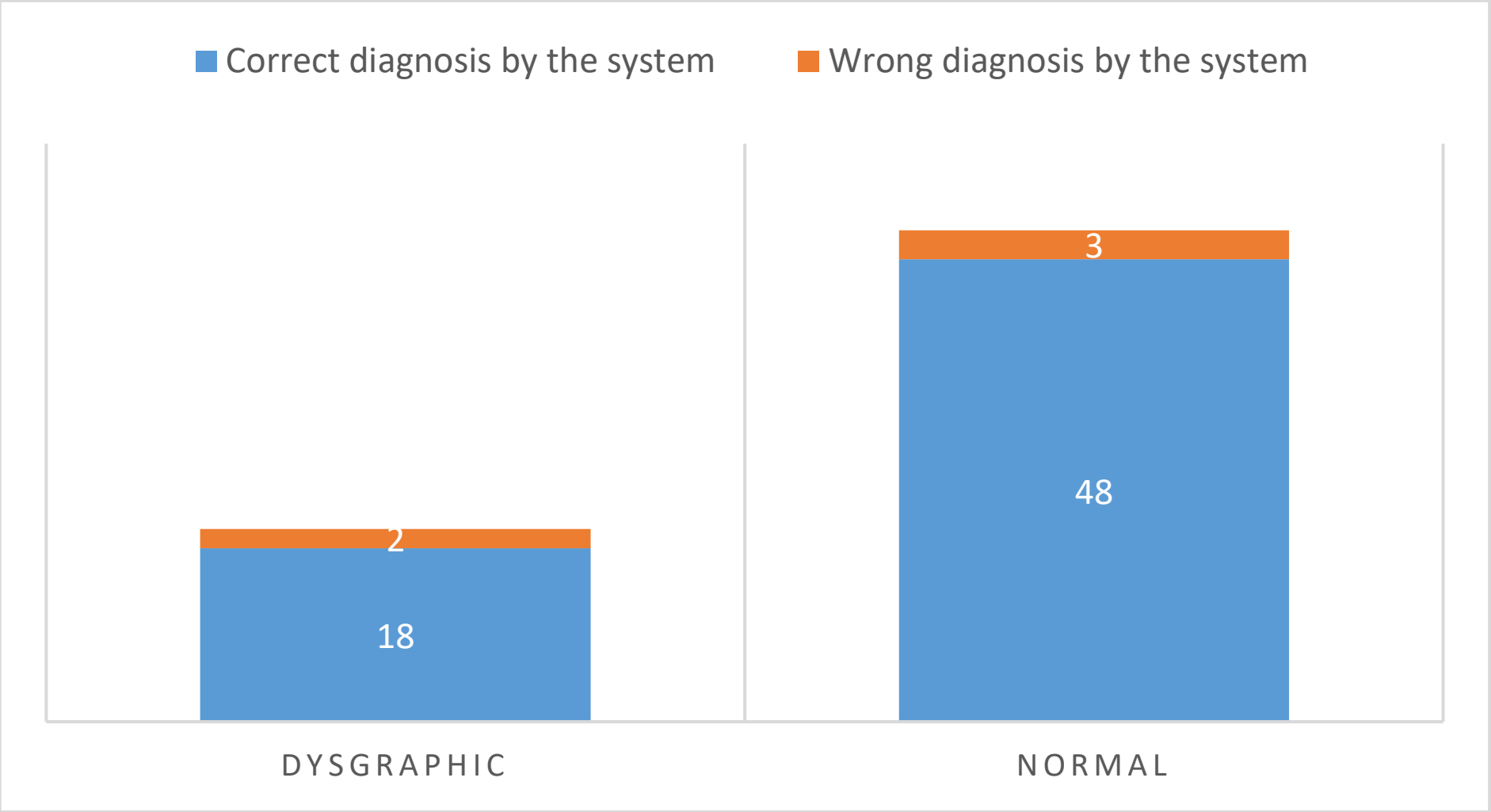

*Table 4 The results of the system about the people who entered the main game*

Table 5 is the false-true positive and false-true negative table.

| System diagnosis: dysgraphic | System diagnosis: normal | Real conditions |
|---|---|---|
| 3 | 48 | normal |
| 18 | 2 | dysgraphic |

*Table 5 false-true positive and false-true negative*

Therefore, this study has developed a system with the following advantages:

Low cost

High availability

No stress for the child

No dependence on writing ability

This system can be used in pre-primary ages to predict the risk of dysgraphia and then provide therapeutic intervention before entering school.

Considering the three cases of finding inappropriate conditions for the test, all three cases were approved by the coach. The total accuracy is 93.24%.

## Future works

In this research, the effect of children's bilingualism has not been measured. It is suggested to consider this factor in future research.

In this research, the system testing was entrusted to the teachers. Only the real children's label and the system's announcement label were received to ensure participant confidentiality. If the game information was also collected, it could be used for re-training the system (enforcement learning).

Some dysgraphic children differ from normal children not only in visual attention but also in auditory components. It is possible to increase or not increase the accuracy of the system by adding auditory measurement.

## 7. Refrences